\documentclass[modern]{aastex62}
\turnoffediting

\graphicspath{{./}{figures/}}

\shorttitle{Electron beams from magnetic reconnection null points}
\shortauthors{Chen et al.}
\begin{document}

\title{Magnetic Reconnection Null Points as the Origin of Semi-relativistic Electron Beams in a Solar Jet}

\correspondingauthor{Bin Chen}
\email{bin.chen@njit.edu}

\author[0000-0002-0660-3350]{Bin Chen}
 \affiliation{New Jersey Institute of Technology, 323 Martin Luther King Blvd, Newark, NJ 07102, USA}
\author[0000-0003-2872-2614]{Sijie Yu}
 \affiliation{New Jersey Institute of Technology, 323 Martin Luther King Blvd, Newark, NJ 07102, USA}
\author{Marina Battaglia}
 \affiliation{University of Applied Sciences and Arts Northwestern Switzerland, Bahnhofstrasse 6, 5210 Windisch, Switzerland}
\author{Samaiyah Farid}
 \affiliation{Harvard-Smithsonian Center for Astrophysics, 60 Garden St, Cambridge, MA 02138, USA}
 \altaffiliation[Also at ]{Vanderbilt University, 6301 Stevenson Center Ln., Nashville, TN 37235, USA}
\author{Antonia Savcheva}
 \affiliation{Harvard-Smithsonian Center for Astrophysics, 60 Garden St, Cambridge, MA 02138, USA}
\author{Katharine K. Reeves}
 \affiliation{Harvard-Smithsonian Center for Astrophysics, 60 Garden St, Cambridge, MA 02138, USA}
\author{S\"am Krucker}
 \affiliation{University of California, Berkeley, 7 Gauss Way, Berkeley, CA 94720, USA}
 \altaffiliation[Also at ]{University of Applied Sciences and Arts Northwestern Switzerland, Bahnhofstrasse 6, 5210 Windisch, Switzerland}
\author{T. S. Bastian}
 \affiliation{National Radio Astronomy Observatory, 520 Edgemont Rd, Charlottesville, CA 22903, USA}
\author{Fan Guo}
 \affiliation{Los Alamos National Laboratory, Los Alamos, NM 87545, USA}
 \altaffiliation[Also at ]{New Mexico Consortium, 4200 West Jemez Rd, Los Alamos, NM 87544, USA}
\author{Svetlin Tassev}
 \affiliation{Harvard-Smithsonian Center for Astrophysics, 60 Garden St, Cambridge, MA 02138, USA}



\begin{abstract}

Magnetic reconnection, the central engine that powers explosive phenomena throughout the Universe, is also perceived as one of the \edit1{principal} mechanisms for accelerating particles to high energies. Although various signatures of magnetic reconnection have been frequently reported, observational evidence that links particle acceleration directly to the reconnection site has been rare, especially for space plasma environments currently inaccessible to \textit{in situ} measurements. Here we utilize broadband radio dynamic imaging spectroscopy available from the Karl G. Jansky Very Large Array to observe decimetric type III radio bursts in a solar jet with high angular ($\sim$20$''$), spectral ($\sim$1\%), and temporal resolution (50 milliseconds). These observations allow us to derive detailed trajectories of semi-relativistic (tens of keV) electron beams in the low solar corona with unprecedentedly high angular precision ($<0''.65$). We found that each group of electron beams, which corresponds to a cluster of type III bursts with 1--2-second duration, diverges from an extremely compact region ($\sim$600 km$^2$) in the low solar corona. The beam-diverging sites are located behind the erupting jet spire and above the closed arcades, coinciding with the presumed location of magnetic reconnection in the jet eruption picture supported by extreme ultraviolet/X-ray data and magnetic modeling. We interpret each beam-diverging site as a reconnection null point where multitudes of magnetic flux tubes join and reconnect. Our data suggest that the null points likely consist of a high level of density inhomogeneities possibly down to 10-km scales. These results, at least in the present case, strongly favor a \edit1{reconnection-driven electron acceleration scenario}.

\end{abstract}



\newcommand\dml{dm-$\lambda$}

\section{Introduction} \label{sec:intro}

Explosive events on the Sun, thanks to their proximity, serve as an outstanding laboratory to study catastrophic magnetic energy release and particle acceleration processes in great detail, yielding a window into powerful explosions in more extreme astrophysical plasma contexts such as the atmospheres of magnetically active stars \citep{1991ARA&A..29..275H} and possibly, astrophysical jet/accretion disk systems \citep{2010A&A...518A...5D}. During an explosive solar event, a large fraction of electrons can be accelerated to nonthermal energies \citep{1976SoPh...50..153L,2010ApJ...714.1108K}, resulting in a variety of signatures across the electromagnetic spectrum \citep{2017LRSP...14....2B}. The accelerated electrons not only play a crucial role in the dynamics and energetics of explosive solar activities \citep{2017LRSP...14....2B}, but also have the potential of posing threats to the near-Earth space environment, especially for their most energetic form \citep{2006LRSP....3....2S}. 

Fast magnetic reconnection---a plasma process in which magnetic field lines with opposite directions approach each other and reconfigure---is believed to be the \edit1{principal} mechanism responsible for the impulsive energy release in flares and jets. The released magnetic energy is subsequently converted into other forms of energy in accelerated particles, heated plasma, and bulk plasma flows. Observational evidence for magnetic reconnection in flares and jets has been frequently reported in literature. Examples include cusp- or X-shaped loops \citep{1996ApJ...456..840T,2013NatPh...9..489S,2015SoPh..290.2211G,2015NatCo...6E7598S}, current-sheet- or null-point-like features \citep{2003ApJ...596L.251S,2010ApJ...723L..28L, 2010ApJ...722..329S, 2016ApJ...819L...3Z, 2016ApJ...821L..29Z,2017ApJ...835..139S,2018ApJ...854..122W}, plasmoid ejections, plasma inflows/outflows, supra-arcade downflows, and shrinking loops \citep{1995ApJ...451L..83S, 1999ApJ...519L..93M,2008ApJ...675..868R,2009ApJ...697.1569M,2010ApJ...722..329S,2012ApJ...754...13S,2012ApJ...747L..40S,2013ApJ...767..168L,2014ApJ...797L..14T,2015ApJ...807....7R}, which are typically features that resemble the reconnection-associated magnetic geometry or plasma dynamics outlined by the flare-heated thermal plasma.

Radio and hard X-ray (HXR) observations of nonthermal radiation from accelerated electrons provide a complementary view of the high-energy aspects of magnetic reconnection. Previous microwave and HXR observations have revealed nonthermal sources at or above the top of retracted magnetic arcades, implying the presence of accelerated electrons in the close vicinity of magnetic reconnection site \citep[e.g.,][]{1994Natur.371..495M,2008A&ARv..16..155K,2012ApJ...754....9G,2013ApJ...767..168L,2014ApJ...787..125N}. However, it remains an open question where and how the electrons are energized in the context of magnetic reconnection (see, e.g., \citealt{2011SSRv..159..357Z} for a review). This is due, in part, to the lack of sensitivity and/or spatiotemporal resolution needed to directly trace the energetic electrons to their origin, making it difficult to determine whether the nonthermal electrons are accelerated \textit{in situ} or transported from elsewhere. Possible sites for electron acceleration include the reconnection outflow region \citep[e.g.,][]{1997ApJ...485..859S, 1998ApJ...495L..67T, 2008ApJ...676..704L, 2013ApJ...767..168L, 2015Sci...350.1238C}, the underlying retracted magnetic arcades \citep{2008ApJ...675.1645F}, and the magnetic reconnection site itself, i.e., the magnetic null point or current sheet where field lines join and reconnect \citep[e.g.,][]{1985ApJ...293..584H,1996ApJ...462..997L,2000A&A...360..715K,2006Natur.443..553D,2010ApJ...714..915O,2012NatPh...8..321E}. Likewise, little is known about the reconnection sites and the drivers of electron acceleration themselves. 

Coherent radio bursts from electron beams propagating along magnetic flux tubes at semi-relativistic speeds ($\sim$0.1--0.5$c$ where $c$ is the speed of light), known as type III radio bursts, serve as an alternative, but unique means for tracing accelerated electrons in a wide range of coronal heights---from the low corona to well into the interplanetary space (see, e.g., \citealt{2014RAA....14..773R} for a review). This is largely owing to the coherent radiation mechanism itself, which allows radio emission from a relatively small population of nonthermal electrons to be readily detected (giving that the conditions for the radiation are satisfied). The bursts are emitted at a frequency close to the fundamental or harmonic electron plasma frequency: $\nu\approx s\nu_{pe}\approx 8980s\sqrt{n_e}$ Hz (where $s=$1 or 2 is the harmonic number), which depends only on the plasma density $n_e$ of the coronal environment that the beams traverse. Since the electron beams propagate very rapidly, they encounter a range of coronal densities within a short time, thus producing radio emission with a rapid change of frequency in time. Radio imaging of these bursts at a wide range of densely sampled frequencies with a sufficiently high temporal cadence can be used to map the detailed trajectory of a propagating electron beams in the corona, providing an excellent means of tracing these beams to their acceleration sites. 

Previous radio imaging of type III bursts (and their variants such as type J and U bursts) at one or several discrete frequencies, particularly from solar-dedicated Culgoora, Clark Lake, and Nan{\c c}ay radioheliographs, as well as general-purpose radio interferometers such as the Very Large Array, has yielded important results on the location and propagation of electron beams in the corona \citep[e.g.,][]{1980A&A....88..203D, 1987SoPh..108..333G, 1992ApJ...391..380A, 2001A&A...371..333P, 2008A&A...486..589K, 2013ApJ...762...60S, 2016ApJ...833...87C}. However, the full capability of radio imaging spectroscopy with spectrograph-like dense spectral sampling and high cadence has not been realized until very recently with the completion of the LOw Frequency Array (LOFAR; \citealt{2013A&A...556A...2V}), the Murchison Widefield Array (MWA; \citealt{2013PASA...30....7T}), the Mingantu Spectral Radioheliograph (MUSER; \citealt{2016IAUS..320..427Y}), and the upgraded Karl G. Jansky Very Large Array (VLA; \citealt{2011ApJ...739L...1P}). Equipped with the new technique, recent studies that utilize LOFAR and MWA observations of type III bursts at metric wavelengths (i.e., $<$300 MHz; LOFAR and MWA operate at 10--240 MHz and 80--240 MHz, respectively) have drawn significant new insights into the propagation of electron beams in the high corona, as well as their temporal and spatial correspondence with flare energy release in the lower corona \citep{2014A&A...568A..67M,2017NatCo...8.1515K,2017A&A...606A.141R,2017ApJ...851..151M,2018A&A...611A..57M,2018NatSR...8.1676C}. However, in order to trace the electron beams to the immediate vicinity of the primary magnetic energy release and electron acceleration site, presumably located in the low corona where the plasma is denser (typically at the order of 10$^{10}$ cm$^{-3}$; \citealt{2002SSRv..101....1A,2008A&ARv..16..155K}), spectral imaging of type III bursts at shorter, decimetric wavelengths (``dm-$\lambda$'' hereafter) is preferred because $\nu_{pe}$, which increases monotonically with $n_e$, falls into this wavelength range. The first work that utilized radio imaging spectroscopy to study \dml\ type III bursts with the VLA has successfully demonstrated the unique power of this technique in tracing electron beams in the low corona \citep{2013ApJ...763L..21C}. However, at that time the VLA was still in its commissioning phase, and the observation was made in its most compact D configuration (maximum baseline length of 1 km) with only part of the array utilized for imaging (17 out of 27 antennas). Hence the array had not reached its full capacity in terms of angular resolution, spectral coverage, imaging dynamic range, and temporal cadence, all of which are important for obtaining the new results reported here.

Here we present VLA observations of \dml\ type III bursts at 1--2 GHz associated with a solar jet, taking advantage of VLA's unique capability of radio spectral imaging at hundreds of spectral channels along with unprecedentedly high angular resolution ($\sim$20$''$) and temporal cadence (50 milliseconds). These observations allow us to derive detailed trajectories of type-III-burst-emitting, semi-relativistic electron beams in the low corona with sub-arcsecond spatial accuracy ($<$0$''$.65, or $<$480 km on the solar disk) and moreover, clearly distinguish multitudes of electron beams generated in an individual energy release event that lasts for only 1--2 seconds. By combining the radio observations with extreme ultraviolet (EUV) imaging, X-ray data, and magnetic modeling, we determine the origin of each group of fast electron beams as a single, extremely compact ($\sim$600 km$^2$) region in the low corona trailing the erupting jet spire, interpreted as the magnetic reconnection null point. We present the VLA radio observations in Section \ref{sec:radio}. Context EUV and X-ray observations of the associated jet event are presented in Section \ref{sec:context}. In Section \ref{sec:model}, we utilize three-dimensional (3D) magnetic modeling to place the radio, EUV, and X-ray observations into a physical picture. We discuss implications of the observations for magnetic reconnection and electron acceleration in Section \ref{sec:discussion} and briefly conclude in Section \ref{sec:conclusion}.

\newpage

\section{VLA Radio Observations}\label{sec:radio}

The VLA recorded a solar jet event on 2014 November 1 around 19:10 UT in a small active region (unnumbered on the day of observation, but later designated as NOAA Active Region 12203 after it had further developed). The array was in its C configuration, for which the longest baseline was 3.4 km. The observation was made by the full 27-element array in two circular polarizations at 50-ms cadence, with 512 spectral channels of 2 MHz bandwidth, covering the 1--2 GHz frequency L band (15--30 cm in wavelength). The angular resolution of the radio images, determined by the full-width at half maximum (FWHM) of the synthesized beam, was $26.7''\times 13.3''$ at 1.2 GHz (and scales proportionally to $1/\nu$) at the time of the observation. All the images are restored with a circular beam of $20''$.

\begin{figure*}[ht]
\includegraphics[width=1.0\textwidth]{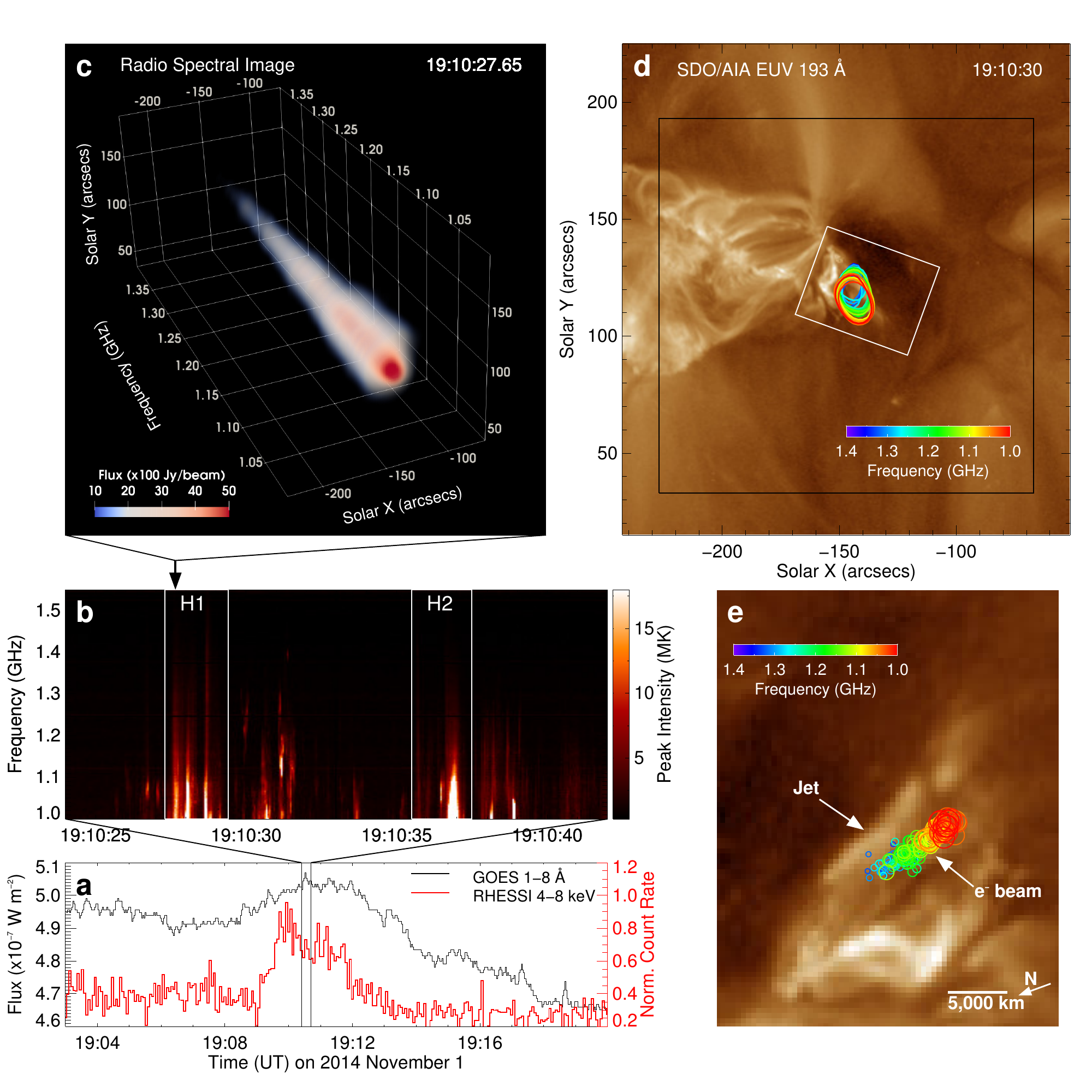}
\caption{\label{fig_tp3} Radio bursts from fast electron beams. (a) Light curves of the solar jet event at two X-ray energies. (b) \edit1{Spatially-resolved intensity  of \dml\ type III radio bursts as a function of frequency and time $I_{\rm pk}(\nu,t)$, or ``vector dynamic spectrum'', recorded by the VLA around the X-ray peak}. (c) Volume rendering of a 3D spectral image cube of an example radio burst made at 200 independent frequency channels within a single 50-ms integration at 19:10:27.65 UT (black arrow indicates the time). (d) Contours of the same type III burst in panel (c) (85\% of the maximum), each of which representing a radio image at a single frequency slice of the 3D spectral imaging cube colored from red to blue in increasing frequency. Background is SDO/AIA 193 \AA\ EUV image (gold; showing coronal plasma at $\sim$1.6 MK). Black box marks the field of view of (c). The ensemble of the image centroids is shown in an enlarged view (e) as colored circles (size scaled by their peak flux), delineating the trajectory of a fast electron beam propagating at \edit1{an apparent speed of $v_{bx}^{\rm app}\gtrsim 0.5c$ in projection}. The display region is marked as the white box in (d) rotated counter-clockwise by 110$^{\circ}$. }
\end{figure*}

\begin{figure}[ht]
\includegraphics[width=1.0\textwidth]{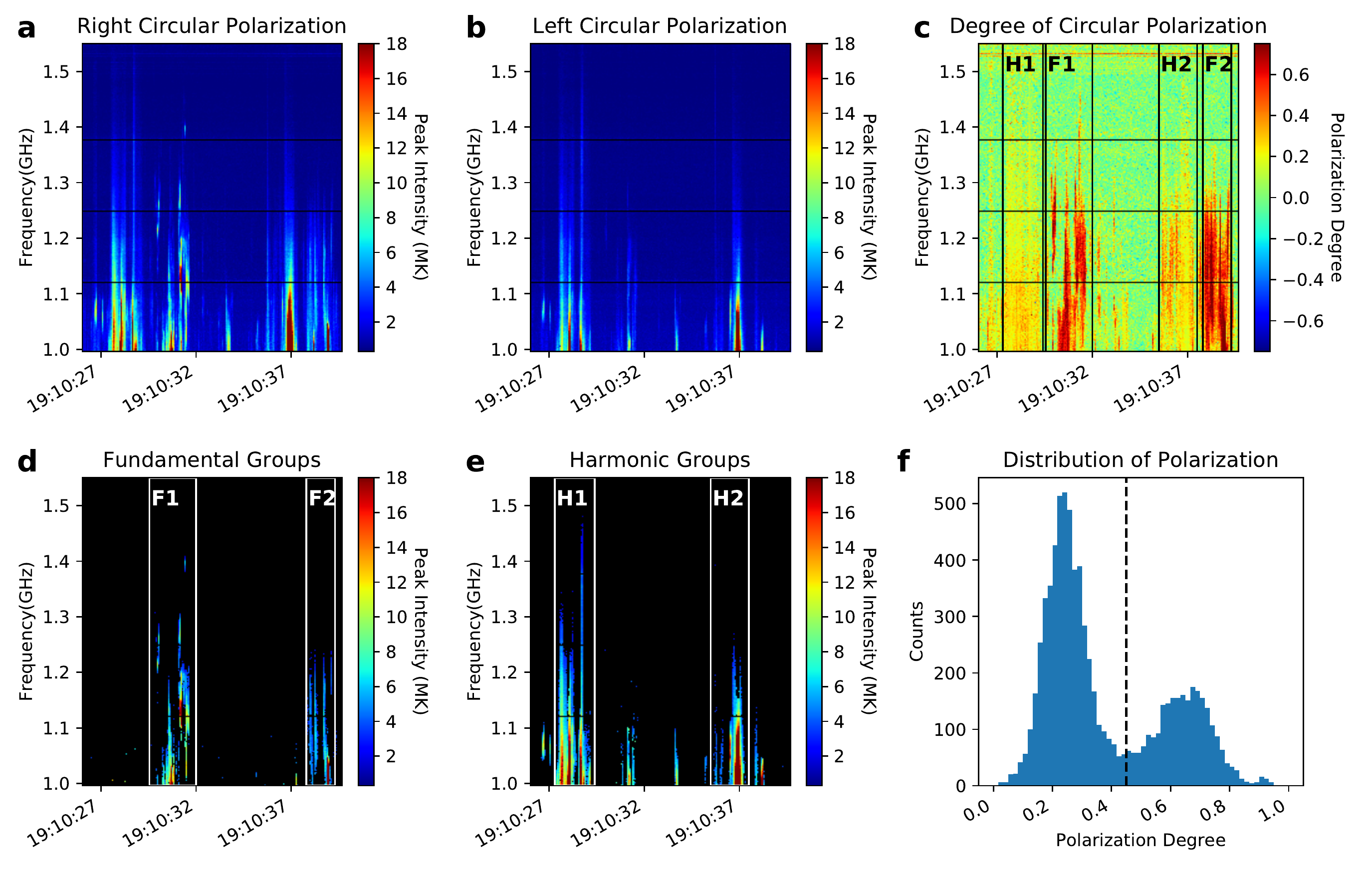}
\caption{\label{fig_tp3_pol} Polarization of the \dml\ type III bursts. (a, b) Dynamic spectrum of the observed \dml\ type III bursts in RCP and LCP, respectively, colored by their \edit1{peak intensity $I_{\rm pk}$ in brightness temperature units}. (c) Same as (a) and (b), but colored by their degree of circular polarization. (d, e) Fundamental (labeled ``F1'' and ``F2'') and harmonic (labeled ``H1'' and ``H2'') burst groups separated by their degree of polarization properties. They correspond respectively to the weakly and strongly polarized groups in the left and right portion of the histogram plot of (f) separated at degree of polarization level of 45\% (vertical dashed line). Only bursts with SNR $>$ 13 are shown in (d-f).} 
\end{figure}

The \dml\ type III bursts were recorded when the X-ray flux reaches its maximum (Fig. \ref{fig_tp3}a-b). The bursts appear in groups, each of which lasts for only 1--2 seconds. Although all the bursts are right-hand-circularly polarized (RCP), they can be clearly divided into two distinctive families in terms of their degree of polarization (DOP; defined as $P=(I_R-I_L)/(I_R+I_L)$, where $I_R$ and $I_L$ are radio intensities in right- and left-hand circular polarization respectively). This is shown in the histogram of Fig. \ref{fig_tp3_pol}f, in which one type III burst family peaks at $P\approx 23\%$ and another at $P\approx 63\%$. The polarization properties of type III radio bursts have been studied extensively \citep{2014RAA....14..773R}. It has been concluded that, in general, type III radio bursts due to fundamental plasma radiation are much more polarized than their harmonic counterpart, although the exact DOP depends on a variety of factors including the magnetic field strength, viewing angle, and detailed wave-mode conversion and propagation processes.

In addition to their different polarization properties, the two type III burst families also clearly differ from each other in their spectrotemporal properties: Bursts in the highly polarized group appear more ``patchy'' than their weakly polarized counterpart (Supplementary Fig. 1a, d). The more chaotic nature for the highly polarized bursts is consistent with the well-known scenario in which fundamental plasma radiation suffers more from propagation effects through the inhomogeneous corona to the observer \citep{1994ApJ...426..774B,2017NatCo...8.1515K}. Therefore, we conclude that the group with higher degree of polarization is likely due to fundamental plasma radiation ($s=1$, or $\nu\approx \nu_{pe}$) and the one with lower degree of polarization is due to harmonic plasma radiation ($s=2$, or $\nu\approx 2\nu_{pe}$). A detailed comparison between the fundamental and harmonic \dml\ type III bursts will be the topic for a future study. Here we choose to focus on the harmonic bursts, because the propagation effects are less significant and the electron beam trajectories derived from the harmonic type III bursts are much better defined.

For each 50-ms time pixel in the radio dynamic spectrum, independent radio images at all the spectral channels are produced using the standard CLEAN image reconstruction technique, resulting in a 3D spectral image cube: Two spatial dimensions in helioprojective X and Y coordinates ($\theta_x$ and $\theta_y$, which are along east-west and south-north direction of the solar disk respectively) and one additional dimension in frequency. An example is shown in Fig. \ref{fig_tp3}c visualized with 3D volume rendering, as well as in Fig. \ref{fig_tp3}d as a series of contours colored from red to blue in increasing frequency. We further obtain the peak intensity $I_{\rm pk}$ in each frequency slice $\nu$ of the spectral image cube and find the corresponding source centroid location ($\theta(\nu)$, $\phi(\nu)$) based on a second-order polynomial fit on nearby pixels. \edit1{The peak intensity values obtained from the interferometric images are originally measured in Jy/beam (as in Fig. \ref{fig_tp3}c), which are subsequently converted to their equivalent brightness temperature values in Kelvin.} The imaging and centroid-fitting processes are repeated for all frequency and time pixels where type III bursts are present, resulting in a four-dimensional (4D) data cube for the burst centroids, i.e., $I_{\rm pk}(\theta, \phi, \nu,t)$. \edit1{One of the many uses of the 4D cube is to construct a ``vector dynamic spectrum'', i.e., the peak intensity variation as a function of frequency and time $I_{\rm pk}(\nu,t)$ obtained within a selected region of interest in the image plane ($\theta, \phi$). Unlike the conventional total-power dynamic spectrum, such a vector dynamic spectrum effectively reduces confusion from other sources if they are also present on the solar disk, thereby reveals the spectrotemporal intensity variation intrinsic to the source of interest itself. An example is shown in Fig. \ref{fig_tp3}b for the \dml\ type III burst source, although the improvement against the total-power dynamic spectrum is minimal because the type IIIs are the only dominating source during the burst period. The power of such technique has been better demonstrated in an earlier study by \citet{2015Sci...350.1238C} based on VLA data: Using vector dynamic spectra made from spatially distinctive regions, several co-existing radio sources including stochastic spike bursts, broadband and narrowband pulsations were separated and identified unambiguously (Fig. S1 in their Supplementary Materials), which were otherwise very difficult to distinguish from the total-power dynamic spectrum. More recently, a similar approach was adopted for the analysis of dynamic spectroscopic imaging data obtained by the MWA, which was discussed in detail by \citet{2017SoPh..292..168M}.}

It has been well known in radio interferometry that the positional uncertainty of the derived source centroid location of a point-like source is well below the angular resolution of the synthesized beam, which is determined by $\sigma\approx\theta_{\rm FWHM}/({\rm SNR}\sqrt{8\ln 2})$, where $\theta_{\rm FWHM}$ is the FWHM beam width and SNR is the ratio of the peak flux to the root-mean-square noise of the image \edit1{\citep{1988ApJ...330..809R,1997PASP..109..166C}}. We select type III burst centroids with SNR $>$ 13 for detailed analysis, which corresponds to a centroid position uncertainty of $\sigma < 0.''65$ at 1.2 GHz, or $<$480 km on the solar disk. Such a high accuracy of the source centroid position is necessary for delineating electron beam trajectories at a length scale of $\sim$1,500--7,500 km for a type III emitting electron beam propagating at 0.1--0.5$c$ within a 50-ms integration. 

Each source centroid at a given frequency ($\theta(\nu)$, $\phi(\nu)$) represents a ``snapshot'' of the radio emission of the electron beam (within the sampling time of the correlator, which is well-below the nominal cadence, 50 ms, determined by the data dump rate), as it reaches a location where the background density $n_e\approx (\nu/8980s)^2$ cm$^{-3}$. As $n_e$ generally decreases with height along an electron-beam-conducting magnetic flux tube (``EB flux tube'' hereafter), the ensemble of the image centroids at decreasing frequencies represents the (projected) trajectory of the electron beam as it propagates upward along a magnetic flux tube. In this event, each electron beam trajectory, obtained within a single 50-ms integration, has a close-to-linear appearance, which delineates the magnetic flux tube along which the beam propagates. An example of such a trajectory is shown in Fig. \ref{fig_tp3}e. Although an individual burst is not resolved in time (i.e., the burst duration $\Delta \tau<50$ ms), the measured length of each trajectory \edit1{$\Delta x$ gives an estimation of the lower limit of the beam's apparent speed in projection $v_{bx}^{\rm app}=\Delta x/\Delta\tau$, which is 0.18--0.53$c$ for all our observed type III bursts. To obtain the actual beam speed $v_b$, however, one needs to know the projection angle between the beam trajectory and the plane of the sky $\alpha$. In the low speed limit ($v_b \ll c$), the relation is simply $v_b=v_{bx}^{\rm app}/\cos\alpha$. However, for electron beams that propagate at a sizable fraction of $c$, the time-of-flight effect and, in some extreme cases, relativistic effect, must be taken into account \citep{1994A&A...286..611P}. Following the approach in \citet{2003A&A...410..307K} that considers the time-of-flight effect (detailed in their Section 5), the actual beam speed is given by $v_b=v_{bx}^{\rm app}/(\cos\alpha+v_{bx}^{\rm app}\sin\alpha/c)$. It is straightforward to obtain the \textit{smallest possible} beam speed that can account for the observed apparent beam speed in projection $v_{b\rm min} = v_{bx}^{\rm app}/(\cos\alpha_{\rm min}+v_{bx}^{\rm app}\sin\alpha_{\rm min}/c)$ where $\alpha_{\rm min}=\arctan(v_{bx}^{\rm app}/c)$, found to be at least 0.18--0.47$c$, or 8.5--68 keV}. Such high-speed, type-III-emitting beam-plasma systems should contain nonthermal electrons with energies well above the value that corresponds to the beam speed $v_b$ \citep{1999SoPh..184..353M}. \edit1{The nonthermal nature of the \dml\ type III bursts is further supported by their high brightness temperature values, which are well above the nominal coronal temperature values of 1--2 MK, sometimes exceeding 20 MK (c.f., Fig. \ref{fig_tp3}b). We note that the intrinsic brightness temperature of the type III sources is likely much higher as they remain unresolved in our images.} 

\begin{figure*}[ht]
\includegraphics[width=1.0\textwidth]{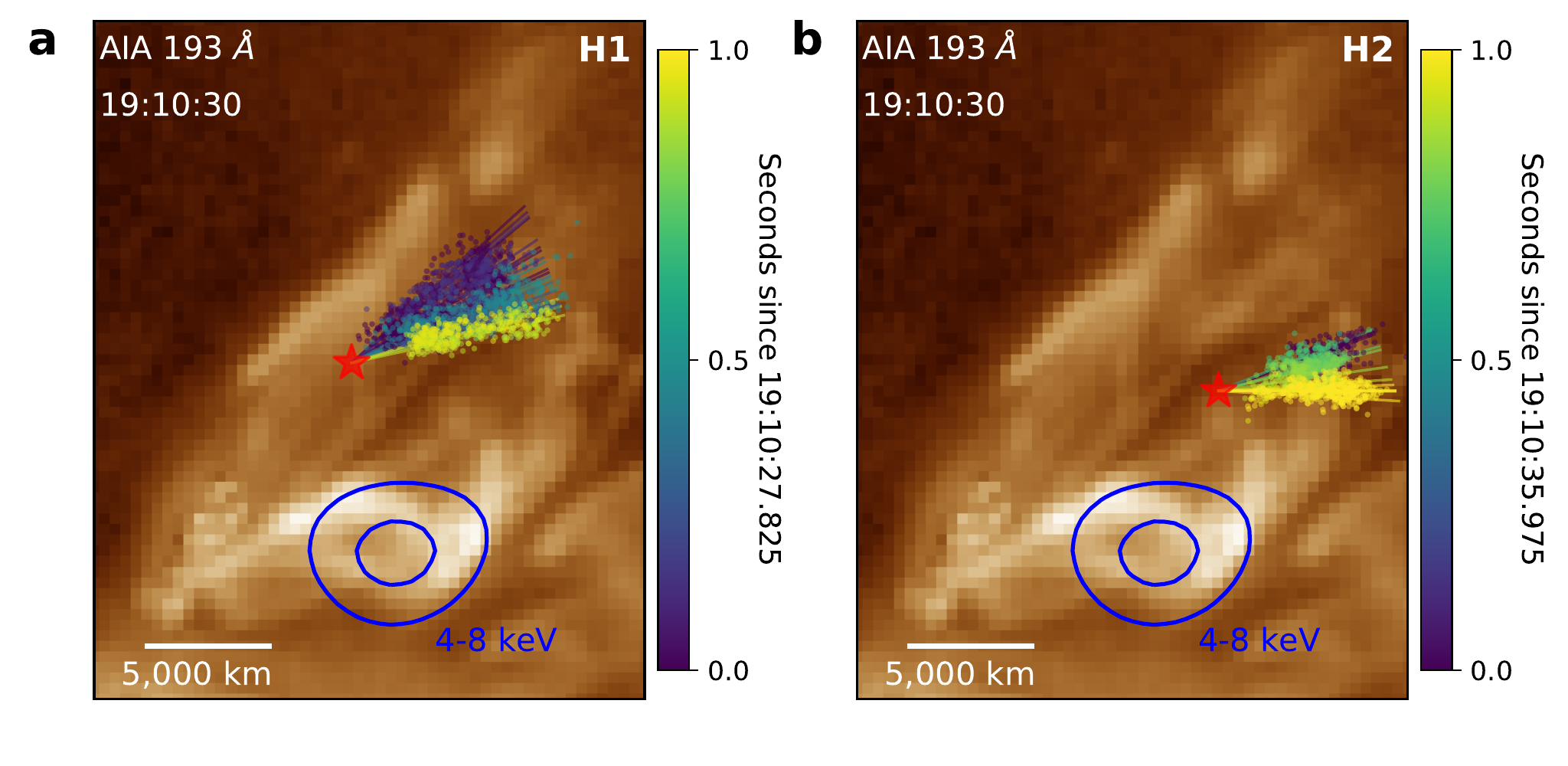}
\caption{\label{fig_cents} Multitudes of electron beams emanating from discrete reconnection sites. Electron beam trajectories for two type III radio burst groups (labeled ``H1'' and ``H2'' in Fig. \ref{fig_tp3}b) are shown in (a) and (b), respectively. Each trajectory, delineated by a series of nearly linearly distributed source centroids (similar to Fig. \ref{fig_tp3}e), is marked with the same color and is outlined by a straight line. Different colors denote beam trajectories at different times. The common reconnection site from which all the electron beams in each burst group originate is denoted by a star symbol. Background is SDO/AIA 193 \AA\ EUV image. Blue contours are RHESSI X-ray source at 4--8 keV (60\% and 90\% of maximum), from hot, retracted arcades at $\sim$8.7 MK. See online Videos 1 and 2 for animation.}
\end{figure*}

The ensemble of all the electron beam trajectories of a given type III burst group, however, displays an evident spread in their position angles on the plane of the sky. In fact, at some times the position angle of the trajectory of one burst can be different from the next one that occurs 50 ms later, whereas at other times a sequential evolution seems to exist (Fig. \ref{fig_cents}; See online videos 1 and 2 for animation). This phenomenon cannot be attributed to the motion of a single EB flux tube. This is because the time scale involved ($<$50 ms) is much shorter than the dynamical time scale of the coronal plasma (at least several seconds for the observed thousands-of-kilometer-scale motion). Instead, each of the observed trajectories delineates the \textit{instantaneous} topology of a distinct magnetic flux tube to which an electron beam gains access. 

Remarkably, for each of the two burst groups, all the electron beam trajectories identified within a group are seen to diverge from an extremely compact, $\sim$600 km$^2$ region in the low corona (two star symbols in Fig. \ref{fig_cents}). In other words, all the beams in a group originate from a common site where a number of different EB flux tubes converge, within a size comparable to the intrinsic cross-section width of non-flaring coronal structures (100s of km) as suggested by some recent studies based on Hi-C data, the highest-resolution EUV imaging observations to date \citep{2013ApJ...772L..19B, 2017ApJ...840....4A}. The two regions are well separated from each other in both time ($\sim$8 s) and space ($\sim$4,500 km), indicating that they are distinct energy release sites. This observation strongly implicates a magnetic reconnection null point or current sheet as the origin of fast electron beams---This is the central location where different magnetic flux tubes are brought in together by reconnection inflows, reconfigure themselves, and release magnetic energy impulsively. Electrons are accelerated during a highly intermittent ($<$50 ms) reconnection event, observed as a fast electron beam escaping from the same reconnection site but along a different flux tube. 

\section{EUV and X-ray Observations}\label{sec:context}

\begin{figure*}[ht]
\includegraphics[width=1.0\textwidth]{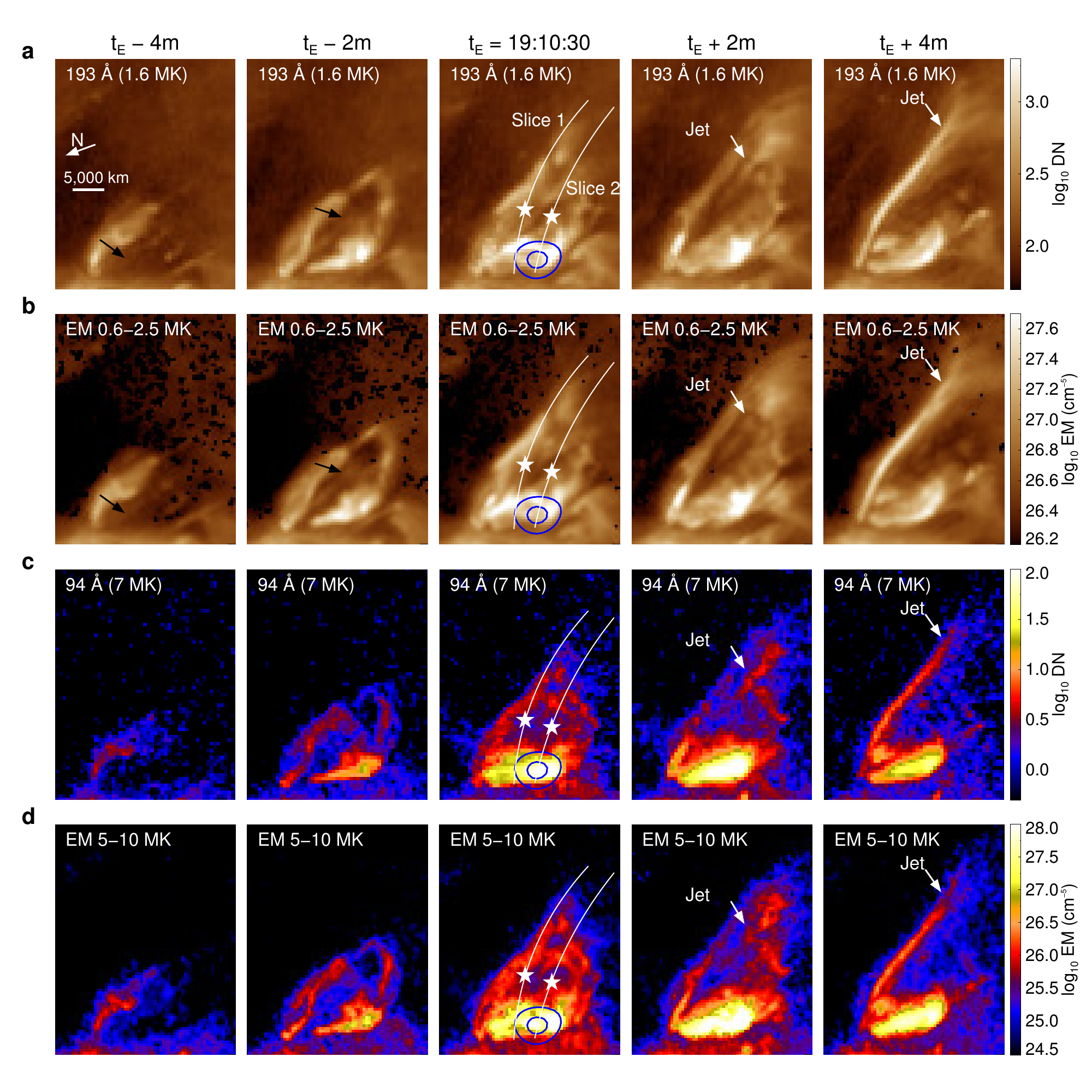}
\caption{\label{fig_jet} EUV observations of the jet eruption. (a, c) Image sequences for AIA EUV 193 \AA\ and 94 \AA, which have peak temperature response at 1.6 MK and 7 MK, respectively. (b, d) Image sequences for the reconstructed column emission measure integrated over 0.6--2 MK and 5--10 MK. The third panels from left (around $t_E=$ 19:10:30 UT) correspond to the time of the \dml\ type III bursts. Star symbols denote the two sites where electron beams originate (c.f., Fig. \ref{fig_cents}). Blue contours are RHESSI X-ray emission from the hot retracted arcades. Black arrows in (a) and (b) indicate the rising filament associated with the jet eruption, visible as a dark feature. The two white curves mark the slices used for obtaining the time-distance plots in Fig. \ref{fig_jet_stackplt}. See online Video 3 for animation.}
\end{figure*}

The Atmospheric Imaging Assembly (AIA) aboard the Solar Dynamics Observatory (SDO; \citealt{2012SoPh..275...17L}) obtained high-resolution imaging observations of the jet event with $1''.5$ angular resolution ($0''.6$ pixel size) at a 12-s cadence in multiple (E)UV passbands. Fig. \ref{fig_jet}a and c show a series of AIA EUV images at 193 \AA\  and 94 \AA, which are sensitive to warm ($\sim$1.6 MK) and hot ($\sim$7 MK) coronal plasma, respectively \citep{2010A&A...521A..21O}. The event starts from a slow rise of a dark filament (indicated by black arrows in Fig. \ref{fig_jet}a and b) embedded in a region with mixed magnetic polarity (Fig. \ref{fig_hmi}), followed by a rapid eruption of plasma and the development of a collimated jet at around 19:10 UT (see online Video 3 for an animation). Meanwhile, bright EUV loops form at the base, coinciding with a X-ray source (4--8 keV; blue contours in Figs. \ref{fig_cents} and \ref{fig_jet}) observed by the Reuven Ramaty High Energy Solar Spectroscopic Imager (RHESSI; \citealt{2002SoPh..210....3L}). Spectral analysis of the X-ray source (see Appendix \ref{appen:xray} for details) suggest these low-lying, compact arcades are heated to $>$8 MK, several times hotter than the ambient coronal plasma. The high-temperature nature of the compact arcades is hinted by the 94 \AA\ observations (Fig. \ref{fig_jet}c), which has a peak at $\sim$7 MK in its temperature response function dominated by iron line Fe XVIII \citep{2010A&A...521A..21O}, and is further confirmed by differential emission measure (DEM) analysis based on the multi-band AIA EUV imaging data: The arcades evidently display enhanced column emission measure (EM) values in 5--10 MK ($\xi_C\approx n_e^2\Delta h$, which is a measure of the amount of plasma along column depth $\Delta h$ integrated within a given temperature range; see Appendix \ref{appen:dem} for details on the DEM analysis).

\begin{figure*}[ht]
\includegraphics[width=1.0\textwidth]{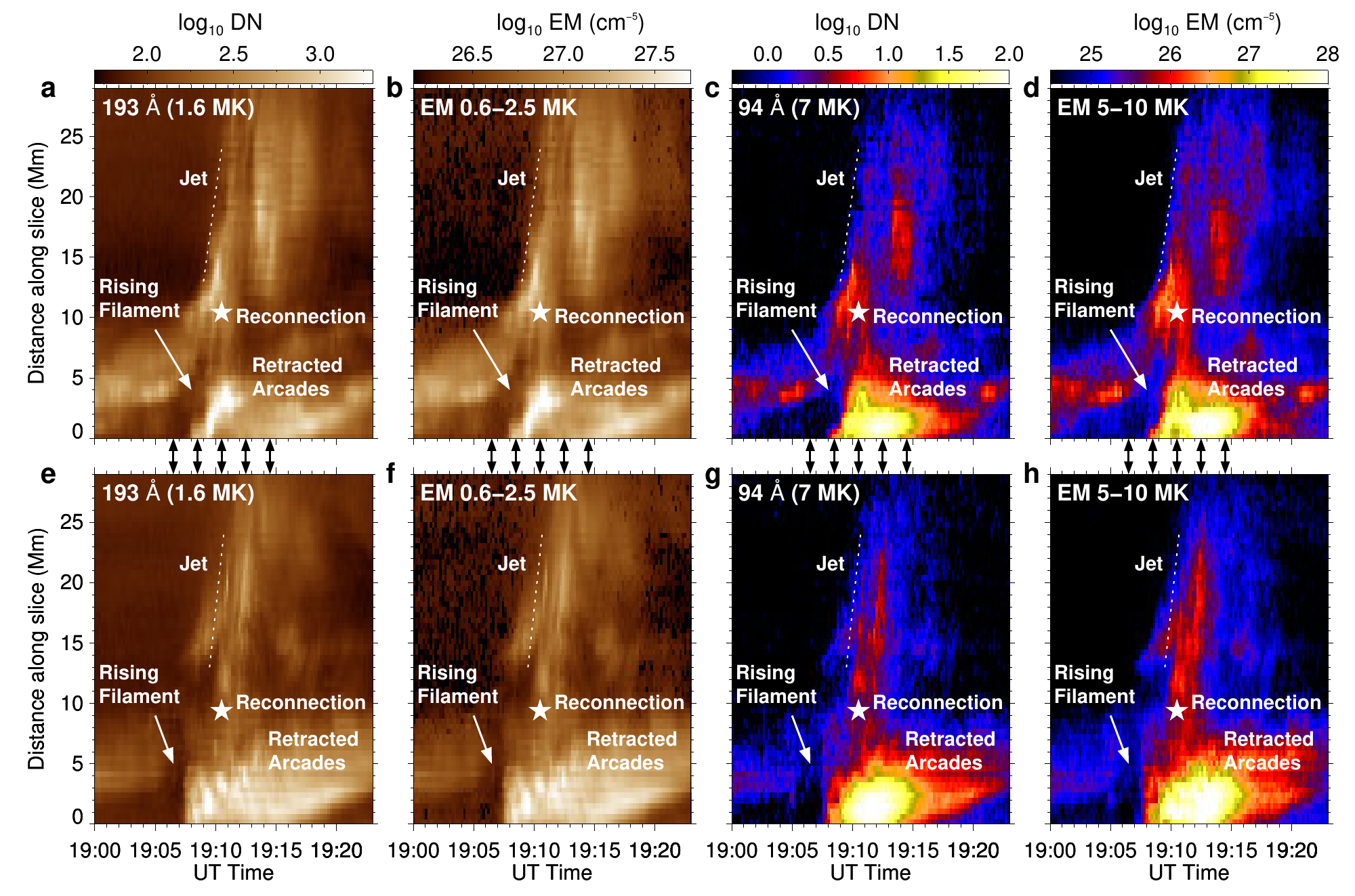}
\caption{\label{fig_jet_stackplt} Time-distance plots of the jet eruption. (a-d) Time-distance plots of SDO/AIA EUV 193 \AA, integrated emission measure in 0.6--2.5 MK, AIA EUV 94 \AA, and integrated emission measure in 5--10 MK, obtained for pixels along Slice 1 in Fig. \ref{fig_jet} that passes through the reconnection site delineated by type III burst group H1. The time and location of the reconnection site is indicated by the star symbol. (e-h) Same as above, but for time-distance plots along Slice 2 in Fig. \ref{fig_jet} that passes the other reconnection site delineated by type III burst group H2. The times of the five snapshot images shown in Fig. \ref{fig_jet} are marked as the double-headed black arrows between the upper and lower panels. }
\end{figure*}

The occurrence of the \dml\ type III bursts during $\sim$19:10:26--19:10:42 UT (c.f., Fig. \ref{fig_tp3}b; note the entire duration of the radio bursts is comparable to a single AIA image frame with a 12-s cadence) coincides very well in time with the eruption phase of the jet (denoted as $t_E$ in Fig. \ref{fig_jet}). Moreover, the two sites from which the electron beams diverge, interpreted above as discrete magnetic reconnection sites, are located just below the erupting jet spire and above the lower-lying arcades where the magnetic reconnection presumably occurs (star symbols in the middle panels of Fig. \ref{fig_jet}). Such a temporal and spatial correspondence is further demonstrated by the time-distance plots in Fig. \ref{fig_jet_stackplt}, which show the EUV intensity as a function of time (horizontal axis) and distance (vertical axis) obtained along two selected slices that pass through the two sites where the electron beams originate (white curves in Fig. \ref{fig_jet}). The erupting jet is seen as the upward-going bright feature with a positive slope (white dashed lines), and the lower-lying, hot arcades are visible below the beam-diverging sites as bright and particularly dense features (c.f. the EM maps in Fig. \ref{fig_jet_stackplt}) that sometimes display negative slopes or downward-contracting motions. It is evident that the \dml\ type III bursts occur when the initially slow rise filament is going through a transition to rapid eruption (horizontal location of the star symbols), and two electron-beam-diverging sites are sandwiched right between the upward-erupting jet and the downward-contracting hot arcades (vertical location of the star symbols). Both the location and timing of the two electron-beam-origin sites agree very well with the presumed site of magnetic reconnection in the physical picture for jet eruption, as will be discussed further in the next subsection.

\section{Magnetic Modeling and Physical Picture}\label{sec:model}

In order to place the radio, EUV, and X-ray observations into the physical context of the jet eruption, we construct a 3D magnetic model of the jet based on photospheric magnetic field data from the Helioseismic and Magnetic Imager (HMI) aboard SDO \citep{2012SoPh..275..207S} using a magnetic flux rope (MFR) insertion method \citep{2004ApJ...612..519V}. This method has been frequently employed to study large-scale erupting and non-erupting phenomena on the Sun that involves a twisted MFR, such as solar flares \citep{2011ApJ...734...53S,2015ApJ...810...96S,2016ApJ...817...43S,2016A&A...591A.141J,2016NatCo...711837X}, sigmoids in solar active regions \citep{2009ApJ...703.1766S,2012ApJ...759..105S}, and quiescent filament channels or coronal cavities \citep{2004ApJ...612..519V,2012ApJ...757..168S}. 

\begin{figure*}[ht]
\plotone{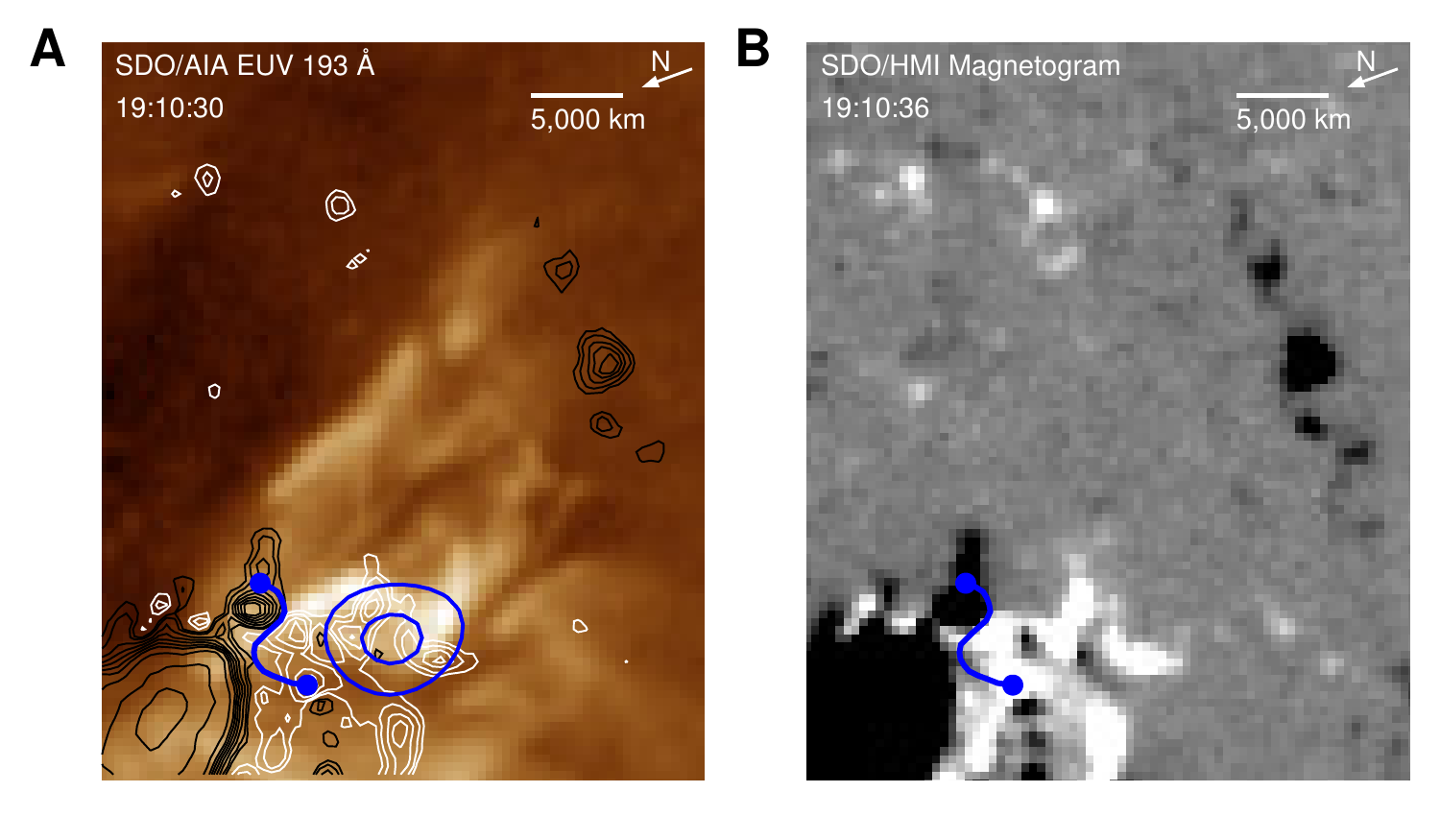}
\caption{Photospheric magnetogram and EUV jet. (a) Line-of-sight magnetic field strength at the solar photospheric level, obtained by the Helioseismic and Magnetic Imager (HMI) aboard SDO. Black and white colors denote negative (southward) and positive (northward) magnetic polarities, respectively. (b) The same magnetogram overlaid on SDO/AIA EUV 193 \AA\ image as black and white contours. RHESSI 4--8 keV X-ray source is shown as blue contours. The jet is rooted at the dominating sunspot with negative magnetic polarity. The blue curve represents the path of the inserted magnetic flux rope at the jet base.}\label{fig_hmi}
\end{figure*}

\begin{figure*}
\includegraphics[width=1.0\textwidth]{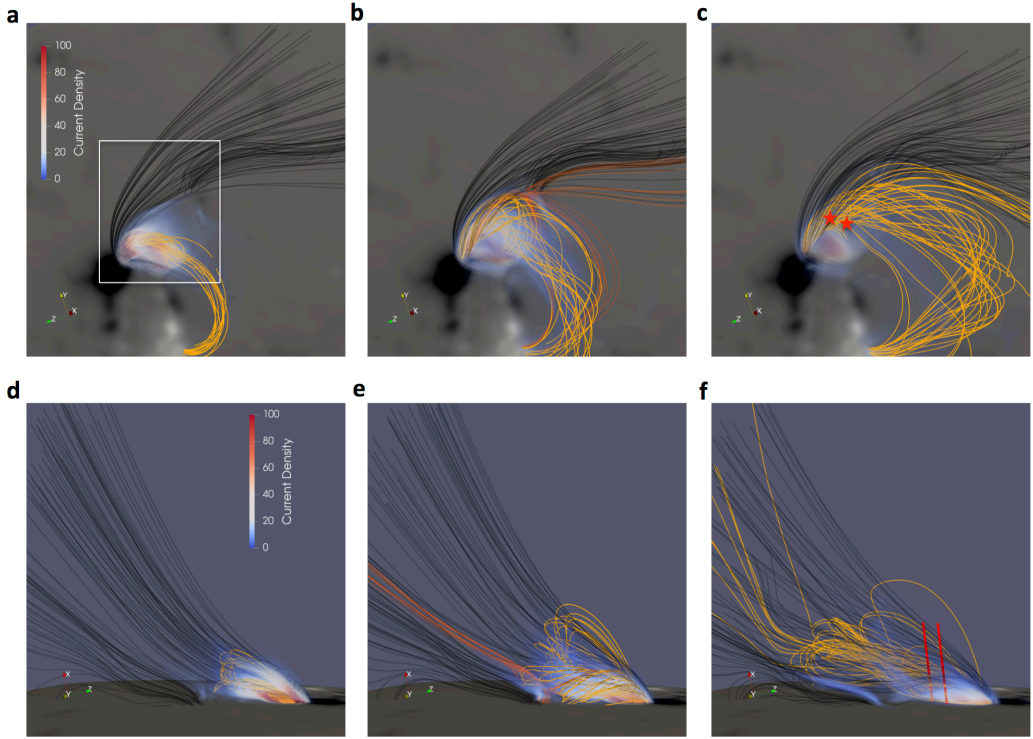}
\caption{\label{fig_model} Three-dimensional magnetic model of the jet eruption. (a-c) Visualization of the magnetic model from the perspective of an Earth-based observer for the pre-eruption, rise, and eruption phase of the jet event. X-, Y-, and Z-axes point toward the LOS, solar west, and solar north, respectively. The field of view displayed in Fig. \ref{fig_cartoon} is shown as a white box. (d-f) Side view of the magnetic model, with the LOS pointing toward the east, or negative Y direction. This is approximately viewing along the axis of the magnetic flux rope. Background grayscale image shows the longitudinal magnetic field at 2 Mm above the photosphere (black and white for negative and positive polarity respectively). Semi-transparent colored shading shows current density in Gauss/Mm (1 Gauss/Mm $\approx$ 0.08 mA/m$^2$) associated with the twisted magnetic flux rope and magnetic reconnection. Black curves represent selected field lines for the ambient open field associated with the sunspots with negative magnetic polarity. Yellow curves represent selected field lines associated with the erupting magnetic flux rope. Orange curves in (b) and (e) depict field lines that interact and reconnect with the ambient field during the flux rope rise phase. The two star symbols in (c) indicate the two sites where the fast electron beams originate (c.f., Fig. \ref{fig_cents}), which are shown in (f) as two straight red lines along the LOS of an Earth-based observer.}
\end{figure*}

\begin{figure}[ht]
\includegraphics[width=1.0\textwidth]{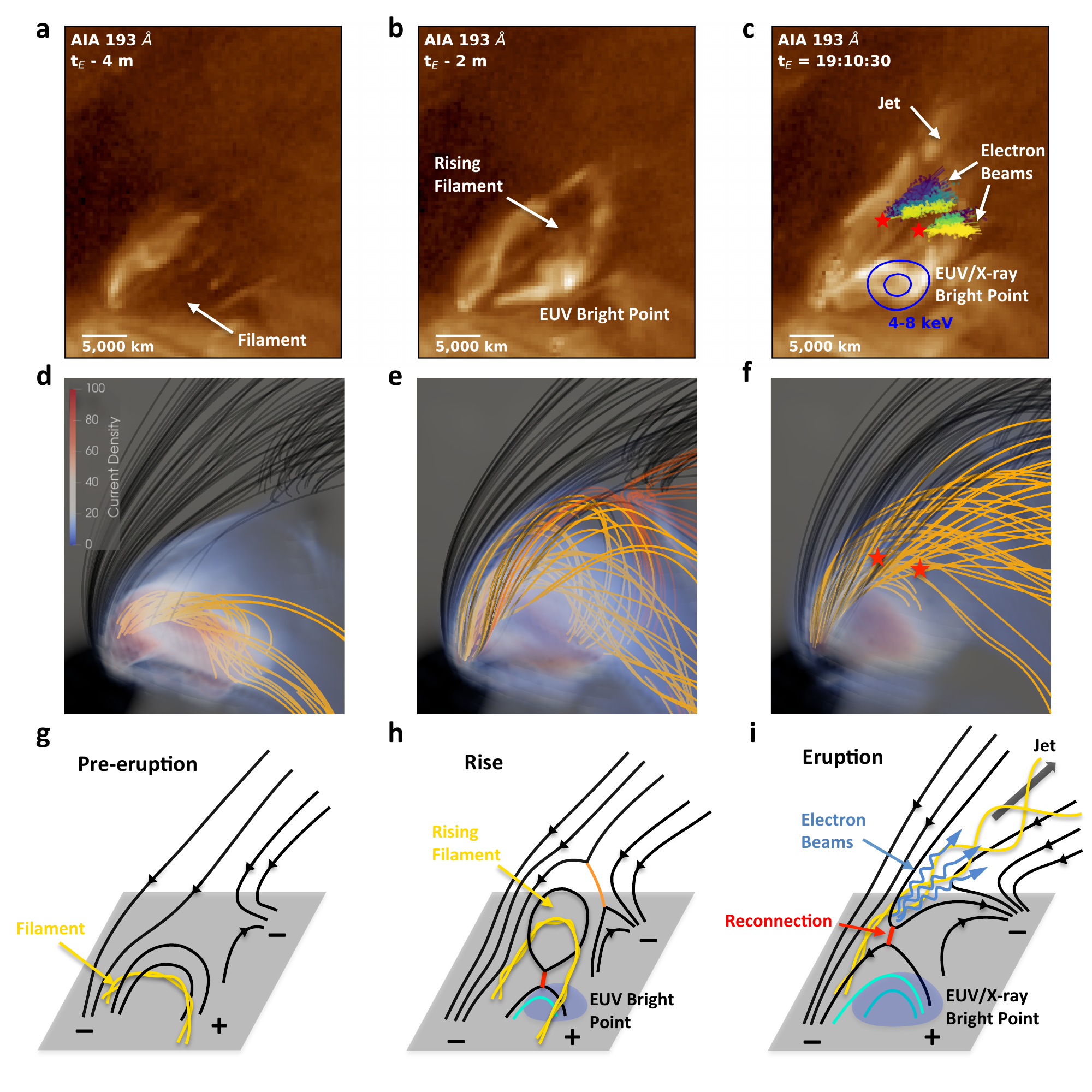}
\caption{\label{fig_cartoon} \textbf{Observations and magnetic modeling of the jet eruption.} Initially an unstable filament visible as a dark feature in EUV, is embedded near the base (a), shown as a twisted magnetic flux rope in the magnetic model (d) and depicted in the schematic (g) as yellow twisted curves. The flux rope slowly rises, pushes against the ambient field (b, e, h), and transfer magnetic flux via slow reconnection (orange curves in e and h). During the eruption phase, fast magnetic reconnection occurs at multiple locations trailing the erupting jet spire (f). Electrons are accelerated from discrete reconnection sites to high speeds and escape along freshly opened field lines (i; only one site is shown for illustration), observed as multitudes of electron beam trajectories emanating from common reconnection sites (c, same as Fig. \ref{fig_cents}). Reconnected and retracted magnetic field lines manifest as compact closed arcades or bright points at the jet base seen in EUV and X-ray. Semi-transparent shading in the model shows current density in Gauss/Mm.}
\end{figure}

Here we, for the first time, adopt the flux rope insertion method to model a small-scale jet event, in light of the ``blowout'' jet scenario that involves the eruption of a small-scale filament or MFR \citep{2010ApJ...720..757M,2015Natur.523..437S}. More detailed descriptions of the MFR insertion method can be found in \citet{2015ApJ...810...96S,2016ApJ...817...43S}. Here we only introduce the procedure briefly. First, we start with a potential field extrapolation using SDO/HMI LOS photospheric magnetogram as the lower boundary condition. Second, a MFR is inserted along the magnetic polarity inversion line close to the path of a dark filament seen in SDO/AIA at the jet base (blue curve in Fig. \ref{fig_hmi}). The magnetic field is then relaxed following a magnetofrictional relaxation treatment \citep{1986ApJ...309..383Y,2000ApJ...539..983V} toward a non-linear force-force free field (NLFFF) configuration. As demonstrated by \citet{2016ApJ...817...43S}, when additional axial flux is introduced for the MFR, the MFR can be rendered unstable. In this case, additional magnetofricion does not lead to equilibrium. Instead, the MFR keeps rising and eventually erupts. Reconnection takes place under the erupting MFR and the configuration keeps evolving. During the process the MFR also interacts with the ambient field and forms a secondary reconnection feature (orange field lines in Figs. \ref{fig_model}b, e and \ref{fig_cartoon}e). The inserted MFR for the NLFFF model has an axial flux of $\Phi_{\rm ax}=1\times 10^{20}$ Mx and a poloidal flux per unit length along the MFR $F_{\rm pol}=-5\times 10^9$ Mx cm$^{-1}$, respectively.

The erupting MFR, represented by selected field lines that thread through a region of enhanced current density in Fig. \ref{fig_model} (yellow curves), as well as its interaction with the ambient magnetic field, are visualized from two different viewing perspectives: One is viewed from an Earth-based observer (top row; same as the radio, EUV, and X-ray observations), and another is viewed from the western side, approximately along the axis of the MFR (bottom row). Although our model does not explicitly simulate the detailed dynamics of the jet eruption, which would otherwise require a full magnetohydrodynamics treatment \citep[e.g.,][]{2017Natur.544..452W}, by comparing to the EUV observations, we are able to identify numerical iterations in the magnetofrictional relaxation process of the magnetic model that best represent the pre-eruption, rise, and eruption phase of the jet (left, middle, and right panels in Fig. \ref{fig_model} respectively). We note that similar practice has been successfully performed for studying larger-scale eruptive flares \citep{2016ApJ...817...43S,2016A&A...591A.141J}. 

The modeling results conform very well with the EUV and X-ray data in the schematic of the blowout jet scenario (Fig. \ref{fig_cartoon}): Initially an unstable filament, or magnetic flux rope, is situated in a region with mixed magnetic polarity (Fig. \ref{fig_hmi}) and ascends gradually (Figs. \ref{fig_model}a,d and \ref{fig_cartoon}a, d, g). It pushes against the overlying field, transfers flux into the side arcades (Figs. \ref{fig_model}b,e and \ref{fig_cartoon}b, e, h), and subsequently erupts. The sudden eruption, observed as an EUV jet, results in fast magnetic reconnection trailing the jet spire (Figs. \ref{fig_model}c,f and \ref{fig_cartoon}c, f, i). Below the reconnection site, the reconnected field lines retract downward and form closed arcades at the base of the jet. They are heated to temperatures several times higher than the ambient coronal plasma, observed as closed arcades or bright point in EUV and X-ray \citep{1996PASJ...48..123S,2010ApJ...720..757M,2015Natur.523..437S}. 

In Sections \ref{sec:radio} and \ref{sec:context} we have shown that type-III-burst-emitting, semi-relativistic electron beams emanate from discrete sites trailing the erupting jet spire and above the retracted lower-lying arcades (star symbols in Figs. \ref{fig_cents}, \ref{fig_jet}, \ref{fig_jet_stackplt}, and \ref{fig_cartoon}). The location of the beam-origin sites coincide with the region where a distribution of enhanced current density is present (Figs. \ref{fig_model}c,f and \ref{fig_cartoon}f). The directions of the electron beams for both burst groups (colored straight lines in Fig. \ref{fig_cartoon}c) are also generally consistent with those of the freshly opened field lines (yellow curves in Fig. \ref{fig_cartoon}f). Although a one-to-one match of the observed EB flux tubes to the model is not achievable due to in part the projection effect, the magnetic modeling results strongly support our conclusion earlier based on radio, EUV, and X-ray observations: The type-III-burst emitting, fast electron beams are likely originated from discrete magnetic reconnection null points induced by the jet eruption.

\section{Discussions} \label{sec:discussion}
Although seemly diverging type-III-emitting electron beams have been reported in a few previous studies \citep{2001A&A...371..333P,2017ApJ...851..151M}, these observations were made at much lower frequencies ($<$300 MHz), thereby the observed type-III-emitting EB flux tubes were located too far away in the upper corona (many tens of thousand kilometers, which is comparable to, or at least a sizable fraction of, the size of the active region) to probe their site of origin in the low corona in detail. In the present study, we are able to precisely pinpoint the beam-origin site to within an region of $\sim$600 km$^2$ where, as discussed earlier, magnetic reconnection most likely occurs. This upper limit of the size of the magnetic reconnection site, determined directly from the radio imaging observations, is comparable to the intrinsic width of non-flaring coronal structures \citep{2013ApJ...772L..19B, 2017ApJ...840....4A} and the spatial scale on which efficient electron acceleration is thought to operate in the low corona \citep{2002SSRv..101....1A}. Moreover, the closest end of the beam trajectories is located only $\Delta d < 1,000$ km away from the beam-origin site (c.f., Fig. \ref{fig_cents}). Previous studies have suggested that a minimum distance $\Delta d_{\rm min}\approx d_{\rm acc}\delta$ is needed for an electron beam to develop sufficient bump-on-tail instability for generating type III bursts, where $d_{\rm acc}$ is the size of the acceleration site (which is the magnetic reconnection site in the present case) and $\delta$ is the power-law index of the electron velocity distribution \citep{2011A&A...529A..66R}. Our observations suggest that the scale of the reconnection/electron-acceleration site should be significantly smaller than 1,000 km and, when escaping from the reconnection site, the electrons have already been accelerated to at least \edit1{tens of keV} along the direction of the reconnecting magnetic flux tubes, probably with a hard energy spectrum (i.e., small $\delta$ values). These observations strongly favor a \edit1{reconnection-driven} particle acceleration mechanism, and place tight observational constraints on particle acceleration theories. For example, if a macroscopic DC electric field is responsible for the electron acceleration, a simple estimate suggest that the corresponding electric field has a lower limit of E $\sim$ 0.1 V/m, well above the typical Dreicer field in the solar corona \citep{1959PhRv..115..238D,1960PhRv..117..329D,1995ApJ...452..451H}. A thorough examination of relevant electron acceleration theories, however, is beyond the scope of this study. 

\begin{figure*}[ht]
\includegraphics[width=1.0\textwidth]{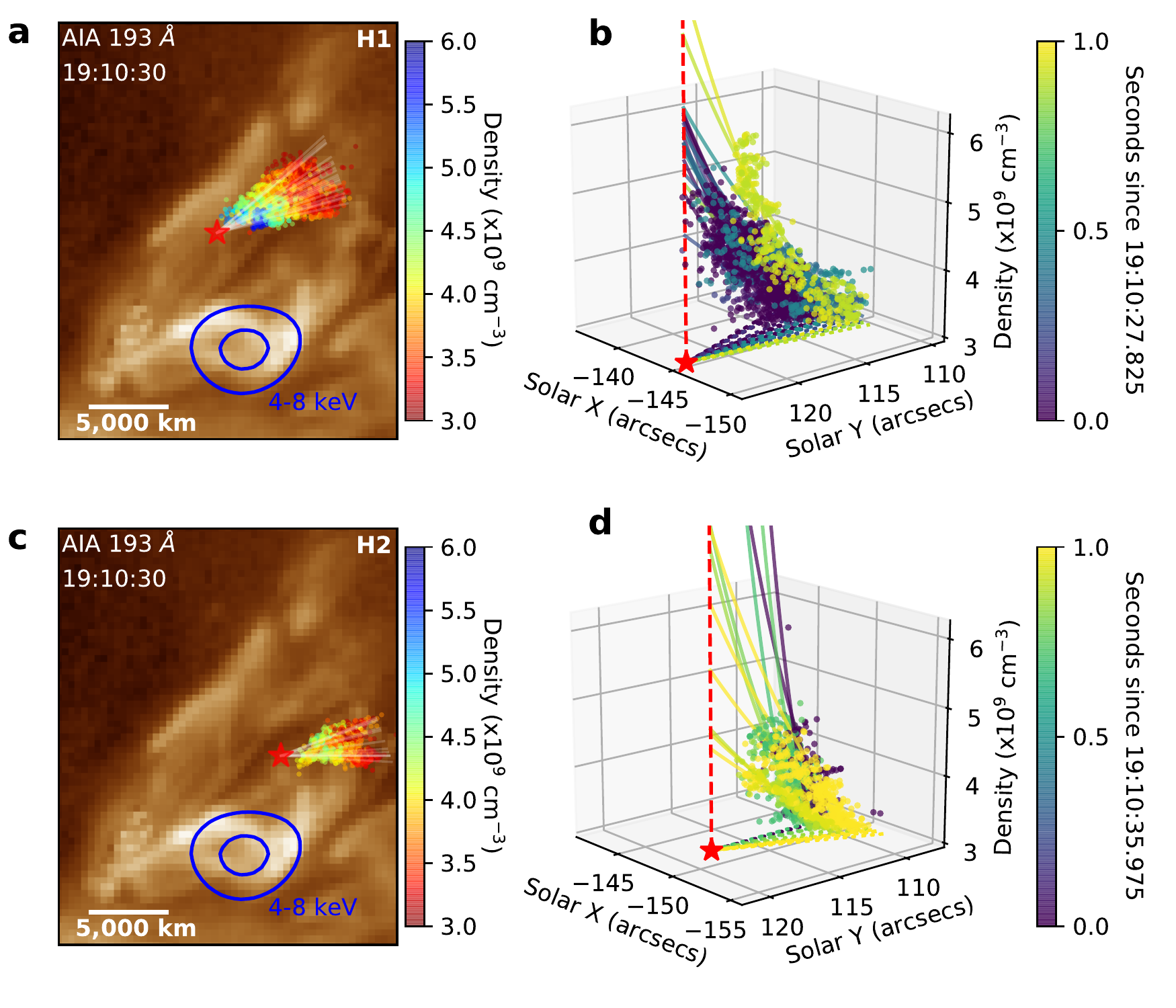}
\caption{\label{fig_cents_density} \textbf{Density distribution along electron beam trajectories}. (a, c) Trajectories for \dml\ type III burst groups H1 and H2 (same as in Fig. \ref{fig_cents}) colored from blue to red in decreasing density. (b, d) Three-dimensional representation of electron density distribution along the electron beam trajectories for burst groups H1 and H2, colored in time. X-Y plane is the plane of the sky in helioprojective coordinates and Z-axis is density. See online Videos 1 and 2 for animation.}
\end{figure*}

We point out that although the upward-erupting jet and downward-contracting arcades, seen in EUV, provide strong evidence that confirms the nature of the electron-beam-origin sites as reconnection null points, the null points themselves are \textit{not} directly detected in the EUV imaging data. As the EUV intensity at each pixel is determined by the column DEM ($d\xi_C/dT\approx d(n_e^2\Delta h)/dT)$ of all the thermal plasma along the line of sight (LOS) convolved with the instrumental response function (Appendix \ref{appen:dem}), \edit1{the ``missing'' EUV counterpart of the null points suggests that the DEM of the EB flux tubes is inadequate for them to be distinguished in EUV images against all the foreground and background structures lying along the same LOS}. \edit1{One possible reason is that} the time elapsed from the production of the electrons, presumably during reconnection, to the moment when they are observed as type III bursts is only $\Delta t = \Delta d/v_b<7$ ms, likely too short for the (essentially \textit{reconnecting}) EB flux tubes to have the chance to be populated with heated plasma that occupies a sufficiently large column depth $\Delta h$. \edit1{Another possibility may be attributed to an insufficient number of nonthermal electrons to induce sizable chromospheric evaporation at the footpoints and subsequently fill the EB flux tubes with heated plasma. The small number of nonthermal electrons is in fact hinted by the nondetection of a nonthermal component in the RHESSI X-ray spectrum above the background (c.f., Fig. \ref{fig_xray}). Yet these electrons are sufficient to produce observable type III radio bursts thanks to the coherent nature of the radiation process.} Similar lines of argument may be applied to earlier spectral imaging studies of type III radio bursts, which appeared to show no EUV loop-like counterpart at the location of EB flux tubes \citep{2001A&A...371..333P,2013ApJ...763L..21C,2017A&A...606A.141R}, although, unlike the present study, the tubes were not determined to the immediate vicinity of the magnetic reconnection site. 

\begin{figure}[ht]
\includegraphics[width=1.0\textwidth]{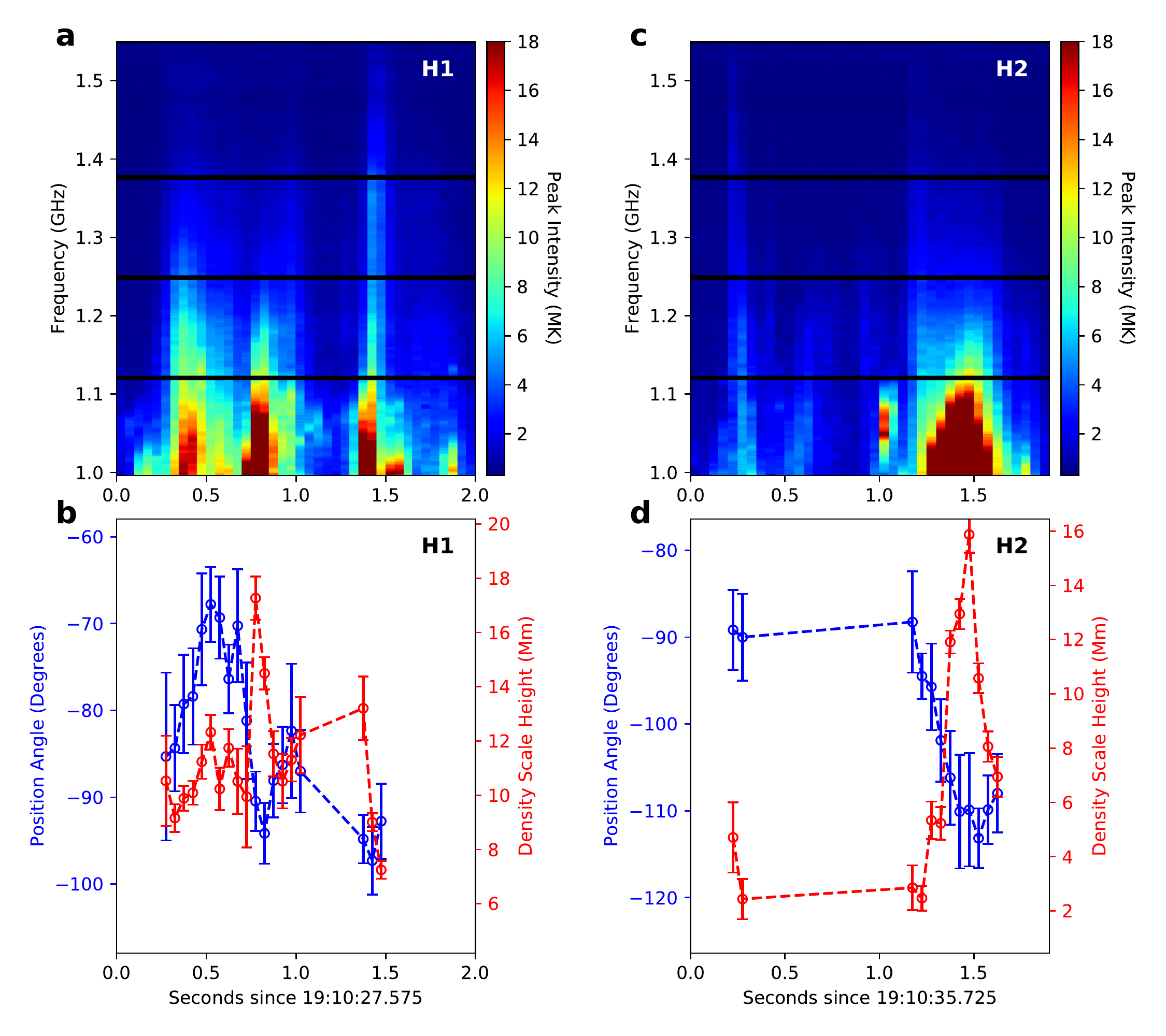}
\caption{\label{fig_pa_ln} Temporal evolution of position angle and density scale height of electron-bream-conducting magnetic flux tubes. The two harmonic \dml\ type III burst groups (``H1'' and ``H2'' labeled in Fig. \ref{fig_tp3}b) are shown in the left and right column, respectively.  (a, c) Radio dynamic spectra of the two harmonic burst groups. (b, d) Evolution of position angle $\phi$  ($\phi=0$ points to solar west; shown in blue) and fitted density scale height in projection ($L^{\rm obs}_{nx}=L^{\rm obs}_n/\tan \alpha$; shown in red) of the type-III-burst-emitting EB flux tubes. }
\end{figure}

Under the interpretation of the observed beam-diverging sites being reconnection null points, our data also provide new opportunity to diagnose physical properties of magnetic flux tubes in the immediate vicinity of the reconnection sites that are probably undergoing active reconnection: As the radio emission frequency depends only on density, one can obtain a density profile $n_e(d)$ along each EB flux tube from the spatial distribution of the radio source frequency $\nu(d)$, where $d$ denotes the projected distance from the reconnection site. The observed density profiles are compared to those expected from a magnetic flux tube under hydrostatic equilibrium: $n_e^{\rm HS}(d)\propto\exp(-d/L^{\rm HS}_{nx})$, where $L^{\rm HS}_{n}=L^{\rm HS}_{nx}\tan\alpha=n_e/(dn_e/dz)=2k_BT/(\mu m_{\rm H} g)\approx 46(T/1~{\rm MK})$ Mm is the density scale height for the hydrostatic case, $m_{\rm H}$ is the mass of hydrogen atom, $\mu\approx 1.27$ is the mean molecular weight for typical coronal conditions, and $\alpha$ is the projection angle \citep{2004psci.book.....A}. We use an exponential function to fit the observed density profile $n_e(d)$ for each derived electron beam trajectory (see Fig. \ref{fig_cents_density}b, d for the fitted curves). All the fitted density profiles are found to be particularly steep, having small values of $L^{\rm obs}_{nx}=L^{\rm obs}_n/\tan\alpha \approx 3$--17 Mm (c.f., Fig. \ref{fig_pa_ln}b, d). Although the projection factor $\tan\alpha$ cannot be determined directly from our observations, \edit1{our magnetic model suggests that the majority of the field lines in the vicinity of the type III sources have inclination angles of $\alpha\lesssim 60^{\circ}$ (c.f., Fig. \ref{fig_model}f)}. Hence the density scale heights of the EB flux tubes are $L^{\rm obs}_n=L^{\rm obs}_{nx} \tan\alpha < 5$--29 Mm, which are at least several times smaller than the hydrostatic values at typical coronal temperatures ($L^{\rm HS}_n\approx 94$ Mm for $T=2$ MK; This could be even larger for flux tubes heated to higher temperatures, which is probably the case for the EB flux tubes near the reconnection site). Such steep density profiles suggest that the EB flux tubes are, at least, far from their hydrostatic state. In other words, they are highly dynamic in nature. One of the possible causes is an upward acceleration of the EB flux tubes, which would result in an increased effective gravity $g_{\rm eff}$ and, in turn, a steeper pressure gradient than the hydrostatic values. This is consistent with the expectation for reconnecting magnetic flux tubes as they remain highly bent and are experiencing strong magnetic tension. Similar to the different position angles of the EB flux tubes, the density profiles also differ from one to another separated by at most 50 ms (Fig. \ref{fig_pa_ln}). It further supports that the EB flux tubes are inherently different in both their orientation and intrinsic dynamical properties. 

The very different density profiles of the EB flux tubes diverging from a single, extremely compact ($<$600 km$^2$) reconnection null point lead to another interesting implication: Since all the EB flux tubes in a burst group converge at the same region, the reconnection null point, in turn, contains much finer spatial structures and is highly inhomogeneous in nature. Fig. \ref{fig_cents_density}b and d show that the extrapolated density values at the reconnection site (i.e., $n_e(d=0)$) vary by at least a factor of two, suggesting a high level of density inhomogeneity of $\delta n_e/n_e>100\%$. One of the possible causes is strong plasma compression at the reconnection site, which may facilitate electron acceleration through a Fermi-type mechanism \citep{2016ApJ...825...55P,2018ApJ...855...80L}.

The observation of tens of fast electron beams produced from an extremely compact region within a very short, $<$50 ms time scale is consistent with the bursty magnetic reconnection scenario \citep{2000A&A...360..715K,2006Natur.443..553D,2002SSRv..101....1A}: The reconnection site is highly inhomogeneous and fragmentary, consisting of many fine dynamic structures such as magnetic islands and fractal currents. Interactions between these fine structures may lead to a burst of electron acceleration. Under this scenario, an upper-limit of the size of the fine structures in the reconnection region is placed to be at the order of $d_{\rm rec}\approx 10$ km. This is based on $d_{\rm rec}\approx u\Delta \tau$, where $u$ is the characteristic speed of the fine structures, taken to be few hundred kilometers per second under typical coronal conditions \citep{2000A&A...360..715K,2002SSRv..101....1A}, and $\Delta \tau < 50$ ms is the duration of an individual acceleration event. Such fine reconnection structures, implied from the spatial-temporal fragmentation of the emanating electron beams, is consistent with the fibrous nature of the reconnection region inferred from radio imaging results (i.e., a $<$600-km$^2$-size reconnection site filled with $>$10 different EB flux tubes), and broadly agrees with earlier implications based on fine temporal and/or spectral structures in radio and HXR light curves (see, e.g., \citealt{2002SSRv..101....1A} for a review). However, here we are able to attribute such the 10-km-scale fine structures directly to individual reconnection null points thanks to our ultra-high-cadence radio imaging data. We note that spatial scales at kilometer-level in the corona is now readily accessible by state-of-the-art kinetic or hybrid numerical simulations \citep{2012NatPh...8..321E,2018ApJ...855...80L}. Observational constraints derived at such fine spatial scales would undoubtedly shed new light on understanding the detailed reconnection-driven particle acceleration processes. 

\section{Conclusion}\label{sec:conclusion}
We have used the VLA to study \dml\ type III bursts in a solar jet with high angular ($\sim$20$''$), spectral ($\sim$1\%), and temporal resolution (50 milliseconds). These observations allow us to distinguish at least ten distinctive, semi-relativistic electron beams associated with each short, 1--2-s duration type III burst group, both spatially and temporally. By mapping detailed trajectories of the electron beams with unprecedentedly high angular precision ($<$$0''.65$), it is revealed that each group of electron beams diverges from an extremely compact region ($\sim$600 km$^2$) located in the low corona behind the erupting jet spire and above the closed arcades. The beam-origin sites coincide very well with the presumed location of magnetic reconnection in the jet eruption picture supported by EUV/X-ray observations and magnetic modeling. Based on the observational evidence and magnetic modeling results, we interpret each of these beam-origin sites as a magnetic reconnection null point. The appearance of the bursts very close ($<$1,000 km) to the reconnection sites strongly favor a reconnection-driven, field-aligned electron acceleration scenario. Our observations suggest that the production of the electron beams are likely associated with a bursty reconnection scenario. Each fast electron beam, which contains energetic electrons of at least \edit1{tens of keV}, is accelerated at or very close to the reconnection null point within tens of milliseconds. From the density profiles along the EB flux tubes, we infer that the reconnection null points likely consist of a high level of density inhomogeneities ($\delta n_e/n_e>100\%$), possibly down to 10-km scales. Our data provide new observational constrains for future theoretical/modeling studies to examine the responsible electron acceleration mechanisms rigorously. Finally, it is intriguing to ask whether this is only an isolated case or otherwise constitutes a general picture for the production of \dml-type-III-burst-emitting electron beams in solar jets. It also remains to be seen whether the picture can be applied to events of much larger scale, e.g., solar flares and coronal mass ejections. Answering these questions calls for further investigations on more \dml\ type III burst events observed with high-resolution, broadband radio dynamic imaging spectroscopy.

\acknowledgments
The NRAO is a facility of the National Science Foundation (NSF) operated
under cooperative agreement by Associated Universities, Inc. We thank D. Gary, G. Fleishman, and L. Glesener for helpful discussions \edit1{and the anonymous referee for his/her constructive comments}. B.C. and S.Y. are supported by NASA grant NNX17AB82G and NSF grants AGS-1654382, AGS-1723436, AST-1735405. A.S. and S.F. are supported by NASA grant NNX15AF43G. K.R. is supported by NASA grant NNX17AB82G. F.G. is supported by NASA grant NNH16AC60I and NSF grant AST-1735414.

\appendix

\section{X-ray observations and spectral analysis}\label{appen:xray}
RHESSI provides imaging and spectroscopy of the Sun in X-rays and γ-rays from 3 keV to 17 MeV \citep{2002SoPh..210....3L}. In this event, the X-ray emission observed by RHESSI is primarily contributed by the retracted, hot post-reconnection arcades located below the reconnection sites, which coincides very well with the hot bright point or compact closed loops seen in EUV (c.f., middle panels in Fig. \ref{fig_jet}c, d at time $t_E=$19:10:30 UT). The 4--8 keV X-ray image shown in Figs. \ref{fig_cents}, \ref{fig_jet}, \ref{fig_cartoon}, and \ref{fig_cents_density} as blue contours is obtained during the jet eruption integrated from 19:10:00 to 19:14:04 UT. The image is reconstructed using the CLEAN algorithm \citep{2002SoPh..210...61H} based on RHESSI measurements from the front segments of detectors 3, 6, and 9. The angular resolution of the RHESSI X-ray image is $\sim$9$''$. 

\begin{figure}[ht]
\includegraphics[width=0.9\textwidth]{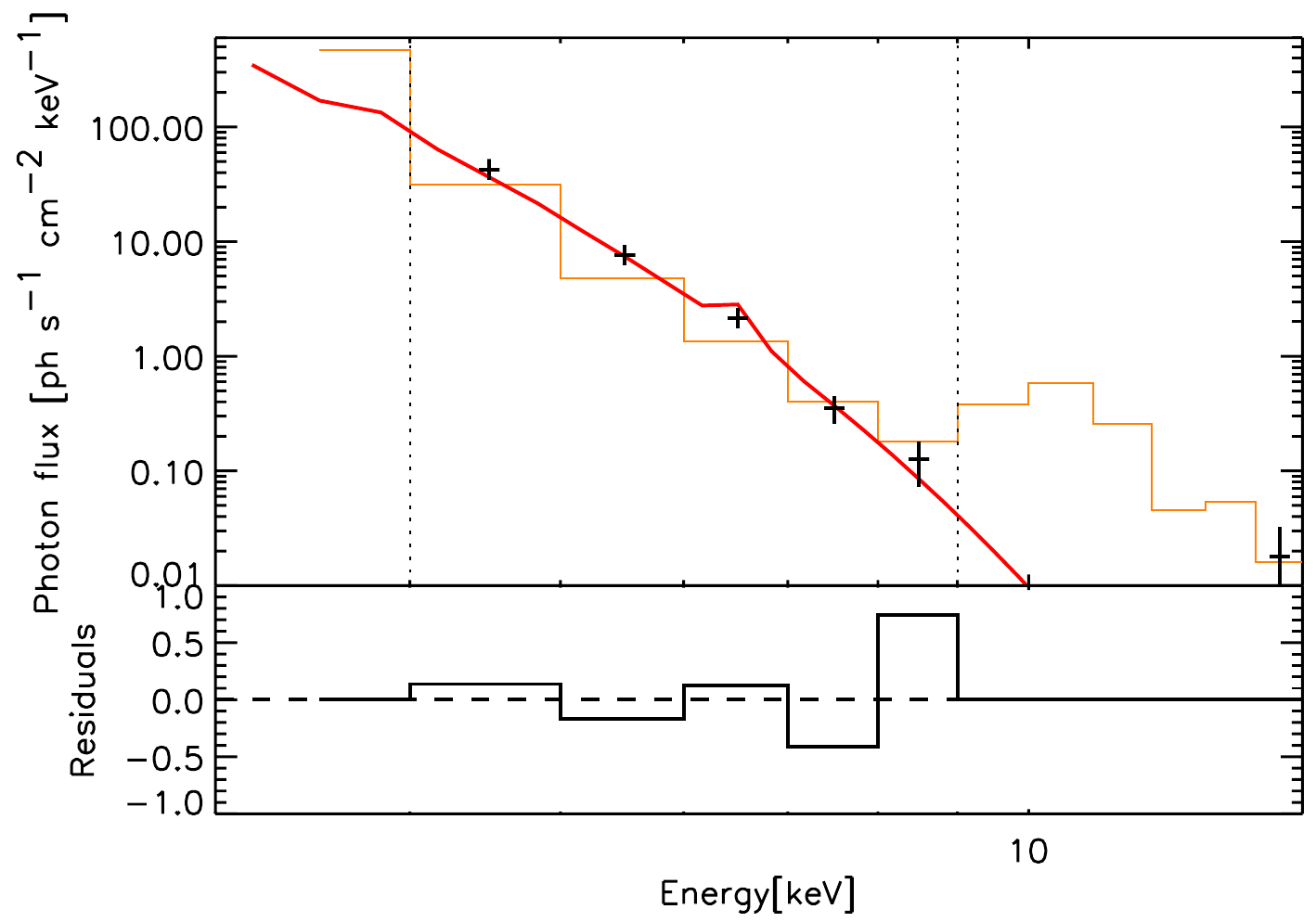}
\caption{RHESSI X-ray spectrum and fit result. Black crosses give the observed X-ray photon flux as a function of energy with uncertainties. Orange histogram indicates the background level at the corresponding energies. Red curve is an isothermal model fitted to the data using abundances in the corona of solar active regions. Residuals of the fit, normalized by \edit1{the one sigma uncertainty of the observed photon flux}, are shown in the lower panel. The fitted emission measure and temperature are  cm$^{-3}$ and 8.7 MK, respectively.}\label{fig_xray}
\end{figure}

The X-ray spectrum is obtained based on measurements from the front segments of RHESSI detector 6 integrated from 19:09:36 to 19:12:00 UT. An isothermal function, calculated using the CHIANTI 7.1 package in the SOLARSOFT IDL distribution \citep{2006ApJS..162..261L}, is used to fit the X-ray spectra from 4 keV to 9 keV (above which the X-ray counts are dominated by background noise). The observed X-ray spectrum and the fitting results are shown in Fig. \ref{fig_xray}. The fitted temperature of the thermal plasma ($T$) is $8.7\pm 0.4$ MK, which agrees well with the EUV results. The volume emission measure is $\xi_V=n^2 V\approx(1.4\pm 0.4)\times 10^{46}$ cm$^{-3}$ (where $V$ is the source volume of the X-ray emitting thermal plasma). Using the FWHM size of the thermal X-ray image ($A\approx 93$ arcsec$^2$), and assuming an equal thickness of the thermal source along the LOS (so $V\approx A^{3/2}$), the number density of the thermal plasma in the flaring loops is estimated to be $n_e=(\xi_V/V)^{1/2}\approx 5.4$--$7.3\times 10^9$ cm$^{-3}$, taking into account the uncertainty in the fitted emission measure. Additional uncertainties are the, essentially unknown, thickness of the X-ray emitting source and its filling factor, i.e. the fraction of the plasma within the volume that is emitting at the observed temperature. For a volume that is one order of magnitude smaller or larger, the density could be up to a factor of three larger or smaller, respectively. The filling factor was assumed to be one. A smaller filling factor would lead to a higher inferred density. Within these uncertainties, this estimated plasma density of the reconnected arcade below the reconnection site is consistent with, but likely higher than, the plasma density of the EB flux tubes above the reconnection site, which is derived from the emission frequencies of the \dml\ type III bursts (3--$7\times 10^9$ cm$^{-3}$).

\section{Differential emission measure analysis}\label{appen:dem}
The observed EUV intensity at each SDO/AIA passband $i$ at a pixel ($x,y$) is due to the distribution of thermal plasma located along the LOS, described by the column differential emission measure (DEM) $d\xi_C(x,y,T)/dT\approx d(n_e^2\Delta h)/dT$ (cm$^{-5}$ K$^{-1}$), where $\Delta h$ is the column depth along the LOS, modulated by the response function of this passband $G_i(T)$: $I_i(x,y)=\int_T [G_i(T)d\xi_C(x,y,T)/dT] dT$. To evaluate the thermal properties of the plasma at the jet site, we adopt a regularized inversion method developed by \citet{2012A&A...539A.146H} to reconstruct the DEM for each pixel based on AIA observations at the six passbands. We then integrate the DEM within selected temperature ranges ($T_{\rm min}$, $T_{\rm max}$) to obtain the integrated column emission measure (EM) $\xi_C=\int_{T_{\rm min}}^{T_{\rm max}} (d\xi_C/dT) dT$ in cm$^{-5}$. Figure \ref{fig_jet} shows image sequences of AIA 193 \AA\ (peak response at 1.6 MK) and 94 \AA\ (the temperature response has one peak at $\sim$7 MK), in comparison with the EM maps integrated in 0.6--2.5 MK and 5--10 MK, respectively. There is a close resemblance between the observed EUV intensities at the two EUV bands and the integrated EM in the corresponding temperature ranges, both spatially and temporally. Hence in the present study we use AIA 193 \AA\ and 94 \AA\ images to represent the ``warm'' background coronal plasma and hot plasma heated by the released energy associated with the jet eruption, respectively. 

\bibliography{chen_2018.bib}

%\begin{thebibliography}{}

%\bibitem[Astropy Collaboration et al.(2013)]{2013A&A...558A..33A} Astropy Collaboration, Robitaille, T.~P., Tollerud, E.~J., et al.\ 2013, \aap, 558, A33 

%\end{thebibliography}



\end{document}